\documentstyle[aps,prc,preprint]{revtex}
\begin{document}

\title{Core-Polarization Contribution to the
Nuclear Anapole Moment} 
\author{N. Auerbach$^{(1)}$ and B. A. Brown$^{(2)}$}
\address{(1) Raymond and Beverly Sackler Faculty of Exact Sciences,
School of Physics and Astronomy, Tel Aviv University,
Ramat Aviv, Tel Aviv 69978, Israel}
\address{(2) NSCL/Cyclotron Laboratory and Department of Physics and Astronomy,
Michigan State University,
East Lansing, MI  48824-1321, USA}
\medskip
\date{\today}
\maketitle
\begin{abstract}
The importance of core contributions to the anapole moment in nuclei
is examined. A model of the core-polarization correction is
presented. The model is based on the coupling of the valence
particles to the spin-dipole $J=1^{-}$ giant resonances of the core.
A shell-model calculation of this correction is presented.
The single-particle moments are calculated with Woods-Saxon
and Skyrme Hartree Fock radial wave functions, and the
general issues associated with nuclear 
configuration mixing are discussed.
\end{abstract}

\newpage

\section{Introduction}

The recent observation of the anapole moment in $^{133}$Cs
[1], has spurred considerable interest in this subject.  As remarked [2], this
is the first observation of a static moment that is due to the
violation of reflection symmetry.  The existence of the anapole moment
was suggested by Vaks and Zeldovich [3] early after the discovery of
parity violation in $\beta $~-decay.

The anapole moment exists in the situation when parity is violated but
time reversal is preserved.  Pioneering calculations of the anapole
moment were done in ref. [4].   It was suggested that the anapole
moment could provide information about the nature of the
nucleon-nucleon (N-N) parity violating force, in particular about the
$\pi $ and $\rho $ exchange contribution to the weak N-N interaction.
One of the immediate applications of the recent $^{133}$Cs
measurement was the attempt to try and deduce the pion-nucleon weak
coupling constant $f_{\pi }$ [2], [5], by comparing the value of the
measured anapole moment with the one calculated using a pure
single-particle model.  This comparison leads to a value for $f_{\pi }$
that exceeded by a factor 4 the value deduced from a hadronic parity
violating measurement in $^{18}F$ [6].  At the present this
controversy still exists.  In this paper we wish to examine
the role of core polarization and calculate its contribution to the
values of the anapole moment.  Some work in this respect has been done in the work
of ref. [7].  Our calculations are analogous to the calculation of
effective charges in nuclei, a concept well rooted in the field of
nuclear structure.

The anapole operator is given by [1], [8]
\begin{equation}\label{1}
\hat{a} = - \pi \int d \vec{r} r^{2} \vec{j} (\vec{r})
\end{equation}
where $\vec{j} (\vec{r})$ is the nuclear electromagnetic current
density.
It has been found [4, 7, 8] that the dominant part of the anapole
operator stems from the spin part of $\vec{j} (\vec{r})$ and is given
by:
\begin{equation}\label{2}
\hat{a}_{s} =   \frac{\pi e \mu}{m} (\vec{r} \times \vec{\sigma })
\end{equation}   
where $\mu$ is the nucleon magnetic moment, $m$ the nucleon mass,
$\vec{\sigma }$ is the nucleon spin operator and $\vec{r}$ is its
coordinate.

For the nucleus we write this part of the anapole operator as

\begin{equation}\label{3}
\hat{a}_{s} = \frac{\pi e}{m} \sum_{i=1}^{A} [\mu + (\mu _{p} - \mu _{n})
 t_{Z}(i)] (\vec{r}_{i} \times \vec{\sigma}_{i})
\end{equation}
where $\mu = \frac{\mu _{p} + \mu _{n}}{2}$, $\mu _{p}$, $\mu _{n}$ are
the proton and neutron magnetic moments in units of nuclear magnitons, 
$t_{Z}$ is the  Z~-~component of
the isospin operator, ($t_{Z}$ for a proton is $+ \frac{1}{2}$, and for
a neutron $- \frac{1}{2}$).   The sum ${\it i}$ is over all nucleons in the
nucleus.  The operator written in vector spherical harmonics
and in terms of a tensor coupled product is:

\begin{equation}\label{4}
\vec{r}_{i} \times \vec{\sigma }_{i} = -i \sqrt{2} 
\sqrt{\frac{4\pi}{3}}r_{i}[Y_{L=1} (\hat{r}_{i})
\otimes \vec{\sigma }]^{\Delta J=1}
\end{equation}
which is the $\Delta J^{\pi}=1^{-}$, L=1, S=1 spin-dipole operator [9].  The anapole
operator in eq. (3) involves therefore, the isoscalar and isovector  
$J=1^{-}$ spin-dipole operators.

The distribution of isovector  spin-dipole strength was studied extensively both
experimentally and theoretically [10] in the eighties and there is
a considerable amount of information about the isovector spin-dipole strength
distribution.  For the isoscalar spin-dipole there is little
information.

The anapole moment is defined as the expectation value of $\hat{a}$.

\begin{equation}\label{5}
a =\, <\psi \mid \hat{a}\mid \psi >_{(J_{z}=J)}
\end{equation}
where $\psi $ is the ground state wave function of the nucleus.  It is
clear that since the ${\it operator}$ $\hat{a}$ is odd under
parity operation and time reversal operations (P-odd, T-odd) for a
spin non-zero nuclear state -${\it  a}$  will be non-zero only if parity mixing
occurs in the wave function $\psi $.
[Note that one does not need time reversal violation in this case
because, the operator $\hat{a}$ contains, unlike for example 
the electric 
dipole, the spin operator $\vec{\sigma }$.  
The anapole ${\it moment}$ therefore is P~-odd and T~-even].

\section{General Formalism}

\subsection{The Single-Particle Contribution}

Let us consider a nucleus with a particle occupying an orbit $j_{+}$
with positive parity.  (The consideration that follow can be made by
starting with a negative parity state $j_{-}$ and interchanging simply
the + and - indices).
The ground state in first approximation can be written as:

\begin{equation}\label{6}
\mid \phi _{+} > = \mid 0^{+}  j_{+} >
\end{equation}  

\noindent
($0^{+}$ denotes the ground state spin of the core).
Consider now a negative parity orbit with the same spin $j$ 
but opposite parity and lying above the $j_{+}$ orbit.  We denote this
orbit as $j_{-}$.

\begin{equation}\label{7}
\mid \phi_{-} > =  \mid 0^{+}  j_{-} >
\end{equation}

In general the negative parity $j_{-}$ orbit will be energetically
about $1 \hbar \omega $  above $j_{+}$.
A parity violating force will mix the two and we will have:

\begin{equation}\label{8}
\mid \phi'_{+} > = \mid 0^{+} j_{+} >_{j} + \eta
_{0}  \mid  0^{+} j_{-}  >_{j} 
\end{equation}

\noindent with

\begin{equation}\label{9}
\eta _{0} = \frac{<\phi _{+} \mid W \mid \phi _{-}>}{\epsilon
_{-} - \epsilon _{+}}
\end{equation}

\noindent
Where $W$ is the parity violating interaction and $\epsilon
_{+}$, $\epsilon _{-}$ are the single-particle energies.  We should remark here
that $W$ is the effective parity violating interaction and may
include also some many-body contributions, such as the excitation of
the $0^{-}$ spin-dipole [11]. The anapole moment from this
admixture is:

\begin{equation}\label{10}
a_{sp}^{(part)} = < \phi'_{+} \mid \hat{a} \mid \phi'_{+} > = 
 2 \eta _{0} < 0^{+} j_{+} \mid \hat{a} \mid
0^{+} j_{-} >
\end{equation}  

In addition to the single-particle contribution involving the $j_{-}$ orbit that is above
the given $j_{+}$ orbit, there is the equally important contribution
of the orbit with spin $j$ equal to $j_{+}$ but of negative parity
lying $1 \hbar \omega $  below the orbit $j_{+}$.  
We denote this orbit as $j^{-1}_{-}$,
indicating that it is a hole state.  The ground state configuration 
$\mid 0^{+}j_{+}>$
will mix with the 2p-1h configuration $\mid 0^{+}j^{2}_{+}j^{-1}_{-}>.$

\begin{equation}\label{11}
\mid \tilde{\phi'}_{+} > = \mid 0^{+} j_{+} >_{j} + 
\tilde{\eta}_{0}  \mid  0^{+} j^2_{+} j^{-1}_{-}  > 
\end{equation}

\noindent
The contribution to the anapole of this mixing, we denote as 
$a^{(hole)}_{sp}$ and
it is:

\begin{equation}\label{12}
a^{(hole)}_{sp} = < \tilde{\phi' }_{+} \mid \hat{a} \mid 
\tilde{\phi'}_{+} > = 
 2 \tilde{\eta}_{0} < 0^{+} j_{+} \mid \hat{a} \mid
0^{+} j^2_{+} j^{-1}_{-} >
\end{equation}  

\noindent
As mentioned, the contribution of this term is of the same magnitude as
$a^{(part)}_{sp}$.  The sum of the two:

\begin{equation}\label{13}
a_{sp} = a^{(part)}_{sp} + a^{(hole)}_{sp} 
\end{equation}  

\noindent
will be considered here to represent the single-particle contribution.
Let us now advance
a bit and include in the wave function configurations involving
excitation of the core.

\subsection{The Core-Polarization Model}

Of the many possible types of core excitations let us single out
the components \mbox{$\mid 1^{-}\otimes j^{\prime }_{+} >$} and 
\mbox{$\mid 1^{-} \otimes j_{-}^{\prime } >$}  involving
single-particle states of positive and negative parity coupled to the
spin-dipole resonances (isoscalar and isovector) to give total spin $j$.
The symbol $\otimes $ denotes, angular momentum coupling.  We will
from now on not write this symbol in order to simplify notations.   
Now we have
\begin{equation}\label{12a}
\mid \psi _{+} > = \alpha \mid 0^{+} j_{+} >_{j}  +
 \sum_{j^{\prime }_{-}} \beta _{j'-} \mid 1^{-} 
j^{\prime }_{-} >_{j}
\end{equation}  
and
\begin{equation}\label{12b}
\mid \psi _{-} > =\bar{\alpha  } \mid 0^{+}j_{-} >_{j} 
+ \sum_{j^{\prime }_{+}} \bar{\beta }_{j'+} 
\mid 1^{-} j^{\prime }_{+} >_{j}
\end{equation} 
In the following we will drop the index $j$ under the kets.
For the sake of simplicity of our presentation let us limit to one
orbit $j^{\prime }$ taking $j_{-}^{\prime } = j_{-}$  and $j^{\prime
}_{+}$ being the next higher positive parity orbit after  $j_{+}$.
An extension to many orbits $j^{\prime }$ is immediate but complicates
matters and notations. We consider therefore:
\begin{equation}\label{13a}
\mid \psi _{+} > = \alpha \mid 0^{+}j_{+} >  + \beta  \mid 1^{-} 
j_{-} >
\end{equation}  
\begin{equation}\label{13b}
\mid \psi _{-} > =\bar{\alpha  } \mid 0^{+} j_{-} >  +
\bar{\beta } \mid 1^{-} 
j^{\prime }_{+} >
\end{equation}
\begin{equation}\label{14a}
\beta = \frac{< 0^{+}j_{+} \mid V_{N} \mid 1^{-} j_{-} >}{\Delta
E_{\beta }}
\end{equation} 
\begin{equation}\label{14b}
\bar{\beta }= \frac{< 0^{+}j_{-} \mid V_{N} \mid 1^{-} j^{\prime }_{+} >}
{\Delta 
E_{\bar{\beta }}}
\end{equation} 
where $V_{N}$ is the nuclear interaction.

The $W$ interaction will mix these two states and the parity
mixed ground state will be:
\begin{equation}\label{15}
\mid \tilde{\psi }_{+} > = \mid \psi _{+} > + \eta \mid \psi _{-} >
\end{equation}
with
\begin{equation}\label{16}
\eta = \frac{ < \psi _{+} \mid W \mid \psi _{-} >}{E_{-} -
E_{+}}
\end{equation} 
Since we expect $\alpha \gg \beta$ we can take $\eta \simeq \eta
_{0}$.   We evaluate $<\tilde{\psi }_{+} \mid \hat{a} \mid \tilde{\psi
}_{+} >$ (dropping the terms quadratic in $\eta _{0}$), and take
 $\alpha \simeq \bar{\alpha } =1$. We find

\begin{equation}\label{17}
a = 2\eta _{0} [< 0^{+}j_{+} \mid \hat{a} \mid
0^{+} j_{-} > + \beta < 0^{+}j_{-} \mid \hat{a} \mid
1^{-}j_{-} > + \bar{\beta } < 1^{-}j^{\prime
}_{+} \mid \hat{a} \mid 0^{+}j_{+} > ]
\end{equation}

At this point we should note that because $\hat{a}$ is a one-body 
operator, the term involving $\bar {\beta }$ will be zero unless
$j^{\prime}_{+} = j_{+}$.  In this case
$\mid \psi _{-} > = \bar{\alpha } \mid 0^{+}  j_{-} > +
\bar{\beta } \mid 1^{-}j_{+} >$.  Then the two
configurations $\mid 0^{+} j_{-} >$ and $\mid 1^{-} j_{+} >$ might be
close in energy (if the spin-dipole is close to its unperturbed
position).  The contribution of this state will be large because of a
large $\bar{\beta }$, however, its contribution will be cancelled by
the nearby orthogonal partner state:
$\mid \psi ^{\prime }_{-} > = \bar{\beta } \mid 0^{+} j_{-}
> - \bar{\alpha } \mid 1^{-} j_{+} >$ .

Therefore:
\begin{equation}\label{18}
a = 2\eta_{0} < 0^{+} j_{+} \mid \hat{a} \mid
0^{+}j_{-} > \times 
[ 1 +  \frac{\beta <0^{+}j_{-} \mid \hat{a} \mid
1^{-} j_{-} > }{< 0^{+} j_{+}
\mid \hat{a} \mid 0^{+} j_{-} >}]
\equiv a_{0} [1 + \chi ]
\end{equation}
where $\chi$ is the core contribution to the anapole moment.

\begin{equation}\label{19}
\chi = \frac{\beta < 0^{+}j_{-} \mid \sum \vec{r}_{i}
\times \vec{\sigma }_{i} \mid 1^{-} j_{-} > }
{<0^{+}j_{+} \mid \sum \vec{r}_{i} \times \vec{\sigma }_i
\mid 0^{+}j_{-} >}
\end{equation}

\subsection{A Simple Estimate}

We now proceed with some simple estimates, treating only the isovector
spin-dipole.
First, $\mid \epsilon_{j+} - \epsilon_{j-} \mid \simeq \hbar \omega $,
\mbox{($\hbar \omega = 41 A^{-1/3}$ MeV)} and $E_{1^{-}} = 1 \hbar \omega +
\Delta V$, where $\Delta V$, is a collective shift due to the
interaction energy of the 1p - 1h states forming the spin-dipole.
The denominator in the expression for $\beta $ is 
approximately equal therefore to \ $\Delta E_{\beta } 
\simeq 2\hbar \omega + \Delta V$.
One can rewrite the expression for $\chi $ as:
\begin{eqnarray}
\chi =
\frac{<0^{+}j_{+} \mid V_{N} \mid 1^{-}j_{-} >
<1^{-}j_{-} \mid \sum \vec{r}_{i} \times \vec{\sigma }_{i}\mid
0^{+}j_{-}>}
{<0^{+}j_{+} \mid \vec{r} \times \vec{\sigma } \mid
0^{+}j_{-} > (2\hbar \omega + \Delta V)}
\end{eqnarray}

\noindent
Proceeding with approximations we may now limit the $\sum \vec{r}_{i}
\times \vec{\sigma }_{i}$ in the matrix elements of the numerator to
the core nucleons and write:
\begin{equation}\label{21}
\chi = \frac{<1^{-} \mid \sum \vec{r}_{i} \times \vec{\sigma }_{i}\mid
0^{+}>}{<j_{+} \mid \vec{r} \times \vec{\sigma } \mid j_{-}>}\cdot
\frac{<0^{+}j_{+}\mid V_{N}\mid 1^{-}j_{-}>}{2\hbar \omega + \Delta V}
\end{equation}

We first note that the ratio 
\begin{equation}\label{22}
\mid \frac{<1^{-} \mid \sum \vec{r}_{i} \times \vec{\sigma }_{i} \mid
0^{+}>}{<j + \mid \vec{r} \times \vec{\sigma } \mid j_{-}>} \mid \simeq
\sqrt {N}
\end{equation}
could be a quite large number depending on the collectivity of the
spin-dipole giant resonance.  The symbol  $N$  stands here for the
effective number of particles that contribute to the collective
spin-dipole.  This number could be a source of enhancement of the core
contribution to ${\it  a}$.  Let us now estimate the other quantities appearing
in eq. (22).

The values of the matrix elements $< 0^{+}j_{+} \mid V_{N}
\mid 1^{-}j_{-} >$, or $<0^{+} j_{-} \mid V_{N} \mid 1^{-}
j^{\prime }_{+} >$ can be estimated from the value of the collective
shift $\Delta V = < 1^{-} \mid V_{N} \mid 1^{-} >$. (See
for example the discussion of particle + core coupling models in ref.
[12]).
{\it On the  average} the values of the above particle + core
coupling matrix elements should be equal to $\frac{\Delta V}{\sqrt N}$ where
$\sqrt N$ is again the number of particles active in the formation of
the
spin-dipole.  We may write the estimate
\begin{equation}\label{23}
\chi = \frac{ \Delta V}{2\hbar \omega + \Delta V}
\end{equation}
denoting $\lambda = \hbar \omega / \Delta V$
\begin{equation}\label{24}
  \chi = \frac{1}{2\lambda + 1}
\end{equation}
For a large collective shift $(\Delta V = \hbar \omega)$ $\chi $ is large.  
In ref. [9] the $1^{-}$ isovector spin-dipole is found to be 
at an energy $2\hbar \omega $, meaning that 
$\Delta V \simeq 1\hbar \omega $.  In this
case $\chi \simeq \frac{1}{3}$.  We should stress that we are dealing
with small admixtures of the core states.  The admixtures for 
$\mid 1^{-} j >$ that are implied here are less than $1\%$.  It is the
factor $\sqrt {N}$ in the spin-dipole strength that makes the $\chi $
correction sizable.  If the collectivity of the spin-dipole is not high,
one will find $\chi $ to be small.
Our estimates are crude and one must have a more precise evaluation of
the contribution of the core.
In the next section, we describe such calculations.

\section{Shell-Model Calculations}

\subsection{Matrix Elements and Operators}

In this section we describe details of the single-particle and
configuration mixed anapole moment calculations.
The spin-current contribution to the anapole moment is given by

\begin{equation}\label{a1}
a_{s} =\, <\psi \mid \hat{a}_{s}\mid \psi >_{(J_{z}=J)} \,
= \left( \begin{array}{ccc} J & 1 & J  \\ -J & 0 & J \end{array}\right)
<\psi \mid \mid \hat{a}_{s}\mid \mid \psi >,
\end{equation}

\noindent 
  where $  J  $ is the nuclear spin, () is the three-j symbol,
and we use reduced matrix
element convention of Edmonds [14].
In units where $  \hbar =c=1  $, the operator $  \hat{a}_{s}  $ is 
given by

\begin{equation}\label{a2}
\hat{a}_{s} =
\frac{\pi  e}{m}
\displaystyle\sum _{i=1}^{A}
\, \mu _{i} (\vec{r}_{i}\times\, \vec{\sigma}_{i})
= -\frac{i \sqrt{2}\, \pi  e}{m} \,
\sqrt{\frac{4\pi }{3}}\,
\, \hat{a}'_{s},
\end{equation}

\noindent 
  where

\begin{equation}\label{a3}
\hat{a}'_{s} =
\displaystyle\sum _{i=1}^{A}
\mu _{i} r_{i} [Y_{L=1}(\hat{r}_{i}) \otimes \vec{\sigma}_{i}]^{\Delta 
J=1},
\end{equation}

\noindent 
and where $\mu_i$ are the nucleon magnetic moments in units of
nuclear magnetons; $  \mu _{p}=2.79  $ and $  \mu _{n}=-1.91  $.

It is conventional to relate the anapole moment to
a dimensionless constant $  \kappa _{s}  $ defined by:

\begin{equation}\label{a4}
a_{s} = \frac{1}{e} \frac{G}{\sqrt{2}}\, \frac{K \kappa _{s}}{J(J+1)}
<\psi \mid \vec{J}\mid \psi >_{(J_{z}=J)} \,
 = \frac{1}{e} \frac{G}{\sqrt{2}}\, \frac{K \kappa _{s}}{(J+1)},
\end{equation}

\noindent 
  where

\begin{equation}\label{a5}
K=(J+\frac{1}{2})(-1)^{\ell +\frac{1}{2}-j}.
\end{equation}

\noindent 
  The $\ell$ and $  j  $ in the
phase factor are chosen to be those of the dominate
single-particle orbital associated with $\psi$.

The perturbation expansion of the reduced matrix element gives:

\begin{equation}\label{a6}
<\psi \mid \mid \hat{a}_{s}\mid \mid \psi >
=-2\displaystyle\sum _{f} \frac{<\psi \mid \mid \hat{a}_{s}\mid \mid 
\phi _{f}><\phi _{f}\mid \hat{W}\mid \psi >}{\mid \Delta E\mid },
\end{equation}

\noindent 
  where $  \hat{W}  $ is the weak interaction.
For the weak interaction we use the approximation of Eq.\ (7)
of [13]:

\begin{equation}\label{a7}
\hat{W}=-\frac{i}{m}\frac{G}{\sqrt{2}}\,
\displaystyle\sum _{i=1}^{A} g_{i} \frac{\vec{\sigma}_{i}}{2} \cdot 
[\vec{\nabla} _{i}\rho +\rho \vec{\nabla} _{i}]
=-\frac{i}{m}\frac{G}{\sqrt{2}}\, \hat{W}',
\end{equation}

\noindent 
  with

\begin{equation}\label{a8}
\hat{W}'= \displaystyle\sum _{i=1}^{A} g_{i} \frac{\vec{\sigma}_{i}}{2} 
\cdot [\vec{\nabla} _{i}\rho +\rho \vec{\nabla} _{i}],
\end{equation}

\noindent 
  where $  g_{p}  $ and $  g_{n}  $ are dimensionless constants
representing the weak-interaction strength between the
valence protons and neutrons, respectively, and the
nuclear-matter density $\rho$.
The nuclear-matter density is normalized by:

\begin{equation}\label{a9}
\displaystyle\int  \rho (r) d\vec{r} = A
\end{equation}

Finally, we express the dimensionless constant $  \kappa _{s}  $ in 
terms
of the matrix elements of $  \hat{a}'_{s}  $ and
$  \hat{W}'  $:

\begin{equation}\label{a10}
\kappa _{s} = \frac{\pi  e^{2}}{m^{2}} \frac{\sqrt{2}\, (J+1)}{K}
\left( \begin{array}{ccc} J & 1 & J  \\ -J & 0 & J \end{array}\right)
\sqrt{\frac{4\pi }{3}}\, \,
\, 2 \displaystyle\sum _{f} \frac{<\psi \mid \mid \hat{a}'_{s}\mid \mid 
\phi _{f}><\phi _{f}\mid \hat{W}'\mid \psi >}{\mid \Delta E\mid }
\end{equation}

\noindent 
  Introducing the units of $\hbar$ and $  c  $, the constant in front
becomes $  \frac{\pi  e^{2} \hbar ^{2} c^{2}}{m^{2} c^{4}}  = 0.199  $ 
MeV fm$^{3}$.
The dimension of the $  \hat{a}'_{s}  $ matrix element is fm
and the dimension of the $  \hat{W}'  $ matrix element is MeV$^{-1}$ 
fm$^{-4}$.

In terms of the weak interaction, $\kappa_{s}$ is a linear combination
of the single-particle coupling constants $  g_{p}  $ and $  g_{n}  $
which we will write as

\begin{equation}\label{a11}
\kappa _{s} = g_{p} \kappa _{sp} + g_{n} \kappa _{sn}.
\end{equation}

\noindent 
  The total value of $\kappa_{s}$ will be obtained using
the values [13] $  g_{p}=4.5  $ and $  g_{n}=0  $ based upon the DDH 
``best value"
estimates [15].
The ultimate goal of comparing these type of calculations
to experimental values for $\kappa$ will be to extract empirical
values for $  g_{p}  $ and $  g_{n}  $ and use these to understand
the nucleon-nucleon PNC weak interaction.
The total value of $\kappa$ also involves the smaller 
convection-current contribution (see Eq.\ 67 of [13]). In this
paper we focus only on the nuclear structure properties of the
most important spin-current term $\kappa_{s}$.

\subsection{Single-Particle Terms}

First we consider the single-particle contributions to the intermediate
states in Eq.\ 39.  The cases we consider are
those given in Table 3 of [13]. Specifically we start with
closed shells for $^{132}$Sn and $^{208}$Pb. For $^{133}$Cs we
take a valence 1g$_{7/2}$ proton particle relative to $^{132}$Sn, for
$^{203,205}$Tl
we take a 3s$_{1/2}$ proton hole relative to $^{208}$Pb, for $^{207}$Pb
we take a 3p$_{1/2}$ neutron hole relative to $^{208}$Pb and for
$^{209}$Bi we take a 1h$_{9/2}$ proton particle relative to $^{208}$Pb.
We will discuss here the results based upon
densities and radial wave functions in the
matrix elements of $  \hat{a}'_{s}  $ and
$  W'  $ which were obtained from Hartree-Fock (HF) calculations
based on the SKX Skyrme interaction
of [16]. The HF results will be compared to those from the
Woods-Saxon (WS) potential of [17] for $^{208}$Pb, and the interpolated
parameter set of [18] for $^{132}$Sn.

In Eq.\ 39 we sum over all single-particle states $\phi$.
The $  \Delta E  $ is the single-particle energy difference.
This sum includes two type of terms: (1) the ``hole" term
in which the
nucleons in occupied states are excited up to orbit $\psi$, e.g.
1p$_{1/2}$ and 2p$_{1/2}$ to 3s$_{1/2}$ for $^{207}$Tl, and (2),
the ``particle" term
in which the nucleon in the orbit $\psi$ is excited
into
unoccupied states $\phi$, e.g.
3s$_{1/2}$ to $  n  $p$_{1/2}$ with $  n \geq 3  $
for $^{207}$Tl. In the oscillator limit the
$  \vec{r}  $ matrix element
is nonzero only for the intermediate states which are 1$\hbar\omega$
away; e.g. 2p$_{1/2}$ and 3p$_{1/2}$ for $^{207}$Tl,
and we find that with the
more realistic HF and Woods-Saxon
radial wave functions, that these ``1$\hbar\omega$" terms
are the only important ones. In cases where the unoccupied states are
loosely bound or unbound, their radial wave functions
are calculated by adding an external square well potential
with a radius of 14 fm and a depth
of 20 MeV to the HF potential. This has a negligible effect
on the HF solution, but gives the unbound states a realistic
excitation energy as well as an exponential fall off at
large distance which is similar to the bound state $\psi$.
The results are not sensitive to the exact values for the
depth and radius of the
external potential as long as it is sufficiently deep to bind the
orbits and sufficiently large to not affect the HF bound
state solution. The many-body aspect of the problem introduces
an extra phase factor of $  -1  $ for the ``hole" term in Eq.\ 39.

The HF and WS results are given in Table I. The WS results for the
particle plus hole contributions are very close to the WS results
given by Dmitriev et al. [13]. Furthermore, the HF results are very
similar to WS. Most of the difference between HF and WS is due to the
difference in the single-particle energy denominator of Eq.\ 39.

\subsection{Core-Polarization Correction}

Next we consider the admixture of particle-hole states in $^{208}$Pb.
The calculation is based on the model space shown in Fig.\ 1 of [19]
which for the $^{208}$Pb closed shell has
1g$_{7/2}$, 2d$_{5/2}$, 2d$_{3/2}$, 3s$_{1/2}$ and 1h$_{11/2}$
filled orbits for protons,
1h$_{9/2}$, 2f$_{7/2}$, 2f$_{5/2}$, 3p$_{3/2}$, 3p$_{1/2}$ and 
1i$_{13/2}$
empty orbits for protons,
1h$_{9/2}$, 2f$_{7/2}$, 2f$_{5/2}$, 3p$_{3/2}$, 3p$_{1/2}$ and 
1i$_{13/2}$
filled orbits for neutrons and
1i$_{11/2}$, 2g$_{9/2}$, 2g$_{7/2}$, 3d$_{3/2}$, 4s$_{1/2}$, 2g$_{7/2}$,
3d$_{5/2}$ and 1j$_{15/2}$ empty orbits for neutrons.
We note that this model space includes the necessary orbits
for the ``particle" admixtures in $^{207}$Th and $^{207}$Pb and thus
we will focus our calculations of
the spin-dipole correlation effects on these two nuclei.

The Hosaka G matrix [20] was used for the residual strong interaction,
and the single-particle energies were fixed to reproduce
the experimental single-particle energies as given in Fig.\ 1 of
[19]. The use of the Hosaka G matrix for the $^{208}$Pb
and its comparison with
other G matrix interactions is discussed in [19].

For the 1$^{-}$ states of interest here there are 27 particle-hole
configurations. One of these is spurious and it was removed from the
spectrum by using the method of Glockner and Lawson [21] of applying
a center-of-mass hamiltonian to raise the energy of the spurious
state and remove its effect from the low-lying states of interest.
(Even though this is a large model space, 
there are six dipole excitations missing
space, e.g. 1h$_{11/2}$ to 1i$_{11/2}$ for protons.)
We show in Fig.\ 1, the single-particle dipole and spin-dipole response
for transitions to 1$^{-}$ states in $^{208}$Pb. Specifically, 
the spin-dipole
strength $B(a)$ is given by the reduced matrix element of 
$\hat{a}'_{s}$.
The unperturbed results shown in Fig. 1 were obtained
using the single-particle energies of [19] and with the center-of-mass
hamiltonian, but with no residual interaction. When the Hosaka G matrix
is used, the dipole strength moves from its unperturbed position
of 7$-$8 MeV to to a collective state at 11.7 MeV. The experimental
giant dipole in $^{208}$Pb [12] lies at about 13.5 MeV with an energy
weighted sum-rule strength of about 100$  \%  $ of the classical
sum-rule value of $  14.8 NZ/A = 735  $ e$^{2}$ fm$^{2}$ MeV. 
The total
experimental B(E1) strength is thus about 54 e$^{2}$ fm$^{4}$ and the
total one-particle one-hole (1p-1h) strength in our calculation is 88 
e$^{2}$ fm$^{4}$.
The results for the dipole and spin-dipole strength
functions are shown in Fig.\ 2. We note that the spin-dipole
strength is collective and is pushed up in energy compared to the
single-particle limit of Fig.\ 1, but it is not as collective
as the isovector dipole. The levels for the low-lying
mixed 1p-1h states in $^{208}$Pb obtained with the G matrix interaction
are in excellent agreement with experiment, within
typically 130 keV [22].

Next we recalculate the ``particle" anapole matrix elements for 
$^{207}$Th and
$^{207}$Pb by using ground state wave which include mixing from the
1p-1h states in $^{208}$Pb.
For example for $^{207}$Th, this admixture
consists of the 3p$_{1/2}$ particle coupled to all of the 1p-1h states.
Since this is a ``2$\hbar\omega$" admixture the spurious center-of-mass
motion can only
be removed approximately.
Its strength was chosen so that the
mostly nonspurious admixed states were stabilized around 8-20 MeV
excitation energy above the single-particle
ground state, whereas the mostly
spurious state was pushed to about 100 MeV excitation energy.
If the spurious state is not removed it comes at a very low
excitation energy and mixes strongly with the single-particle
ground state. On the other hand if the center-of-mass hamiltonian
is too strong, all states are moved up too high (because of their
small but non-zero spurious component) and these is no mixing
of the non-spurious components of interest
with the single-particle state.

The results are given in Table II next to the column labeled
``HF part$+$CP". It turns out the the effect of the core-polarization
is rather small, resulting in about a 10$  \%  $ reduction
for $^{207}$Th and a 5$  \%  $ enhancement for $^{207}$Pb. 
These core-polarization corrections arise from both the
$\hat{a}'_{s}$ and $\hat{W}'$ matrix elements in Eq. 39, but
the dominant effect is on the $\hat{a}'_{s}$ term.
The core-polarization admixtures in the ground states is
only about 1$\%$, yet the effect is rather significant. 
These calculations indicate that the core-polarization is
not too large, but not negligible. The calculations for the
core-polarization contributions might be 
expanded in future work by using a perturbation approach.

\subsection{Configuration Mixing and Comparison to Experiment}

The final step for the anapole moment calculations will be to
go from the ``single-particle" nuclei around $^{132}$Sn
and $^{208}$Pb to the multi-valence-particle configurations
involved in those nuclei where measurements have been
carried out, in particular for $^{133}$Cs and $^{205}$Th.
The main complication here is that the anapole moment will
consist of a linear combination of diagonal
(e.g. $  <\psi \mid \mid \hat{a}\mid \mid \psi >  $)
and off-diagonal
(e.g. $  <\psi \mid \mid \hat{a}\mid \mid \psi '>  $)
reduced matrix elements within the valence space.

Good shell-model
hamiltonians exist for $^{205}$Th [23]. It is known in this
case [23] that the diagonal matrix element for the 3s$_{1/2}$ orbit
gets reduced from by about 0.80 compared to its value in $^{207}$Th
due to configuration mixing. We have use the HF wave functions to
calculate all of the single-particle anapole matrix elements
involved in the $^{205}$Th ground state. When these are combined with 
the
$  \Delta J=1  $ one-body transition densities for the $^{205}$Th
ground state, the anapole moment comes out be 0.40; close to 
the 3s$_{1/2}$ single-particle value. As mentioned, the diagonal
matrix element is reduced, but the smaller off-diagonal terms
give some enhancement which cancels out the reduction.
The core-polarization correction considered above
would reduce this to about 0.35.
To obtain the final value of the anapole moment we add the
additional (smaller) term from the convection-current contribution
as given in Table 3 of Dmitriev et al. [13] which is about
-0.09. The total calculated anapole moment is thus about 0.24, which 
should be compared with the experiment value of $-0.22 \pm 0.30$ [24].
The agreement is fair given the large experimental error.

The single-particle value for the spin contribution to the anapole moment
for $^{133}$Sb from Table I is 0.29 which, when added to the
convection-current contribution from Table 3 of Dmitriev (-0.05),
gives a total anapole moment of 0.25. This is in fair agreement
with the results of $0.36 \pm 0.06$ deduced in [2] on the
basis of atomic physics considerations
from the experiment on $^{133}$Cs atoms [1]. However, one cannot
make a final comparison between theory and experiment until
one has a calculation for the core-polarization contribution
as well as for the structure of $^{133}$Cs.
A hamiltonian for this mass region has been developed and applied to the
magnetic moment of $^{137}$Cs [25], where one finds that the diagonal
1g$_{7/2}$ term is within a few percent of its value in $^{133}$Sb.
However, for $^{133}$Cs the spherical dimensions involved in valence
space are extremely large (on the order of 10$^{9}$) and one will have
to explore the use of the deformed single-particle model or the
shell-model Monte Carlo method in order to carry out a reliable
calculation.

\section{Conclusions}

We have investigated anapole moments in heavy nuclei in the framework of
shell-model configuration mixing. The single-particle anapole moments
are broken down into their components coming from the weak-interaction
mixing of particle and hole terms. The sum of these terms calculated
with Woods-Saxon radial wave functions are close to the values obtained
by Dmitriev et al. [13]. We have also evaluated these matrix elements
with Skyrme Hartree-Fock radial wave functions and the results
are similar to
those obtained with the Woods-Saxon potential. We discuss the general
principle behind the core-polarization corrections to the anapole
moment. Specific calculations are carried out for $^{208}$Pb with a G
matrix interaction which incorporates realistic collective states for the
spin-dipole excitation. The core-polarization corrections
for the ``particle" contributions in $^{207}$Th and $^{207}$Pb turn out
to be on the order of $10\%$; not very large, but not negligible. The
single-particle matrix elements which we could consider in these
calculations were, however, limited, and it would be desirable to carry
out more extensive calculations along these lines. We have also made a
configuration mixing calculation for the anapole moment of $^{205}$Th
which gave a value close to the previous single-particle estimate and in
fair agreement with experiment.

\vspace{ 12pt}
Acknowledgements: This work was supported in part by the
US-Israel Binational Science Foundation and by NSF grant PHY-9605207.

\newpage
\noindent
Figure Captions:

\vspace{ 12pt}
\noindent
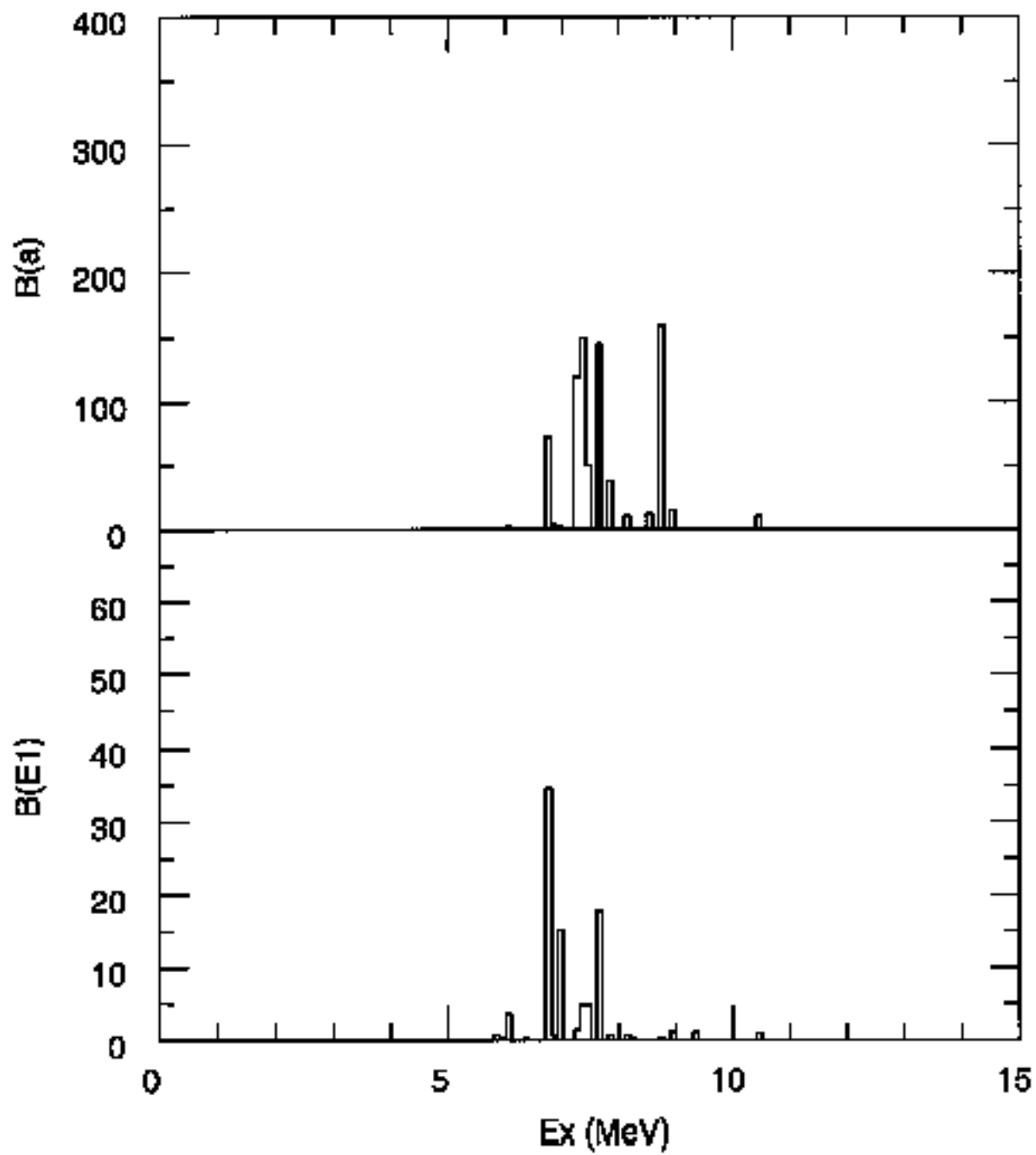
Figure 1: $B(E1)$ and $B(a)$ strength distributions in $^{208}$Pb
obtained with the unperturbed 1p-1h wave functions.

\vspace{ 12pt}
\noindent
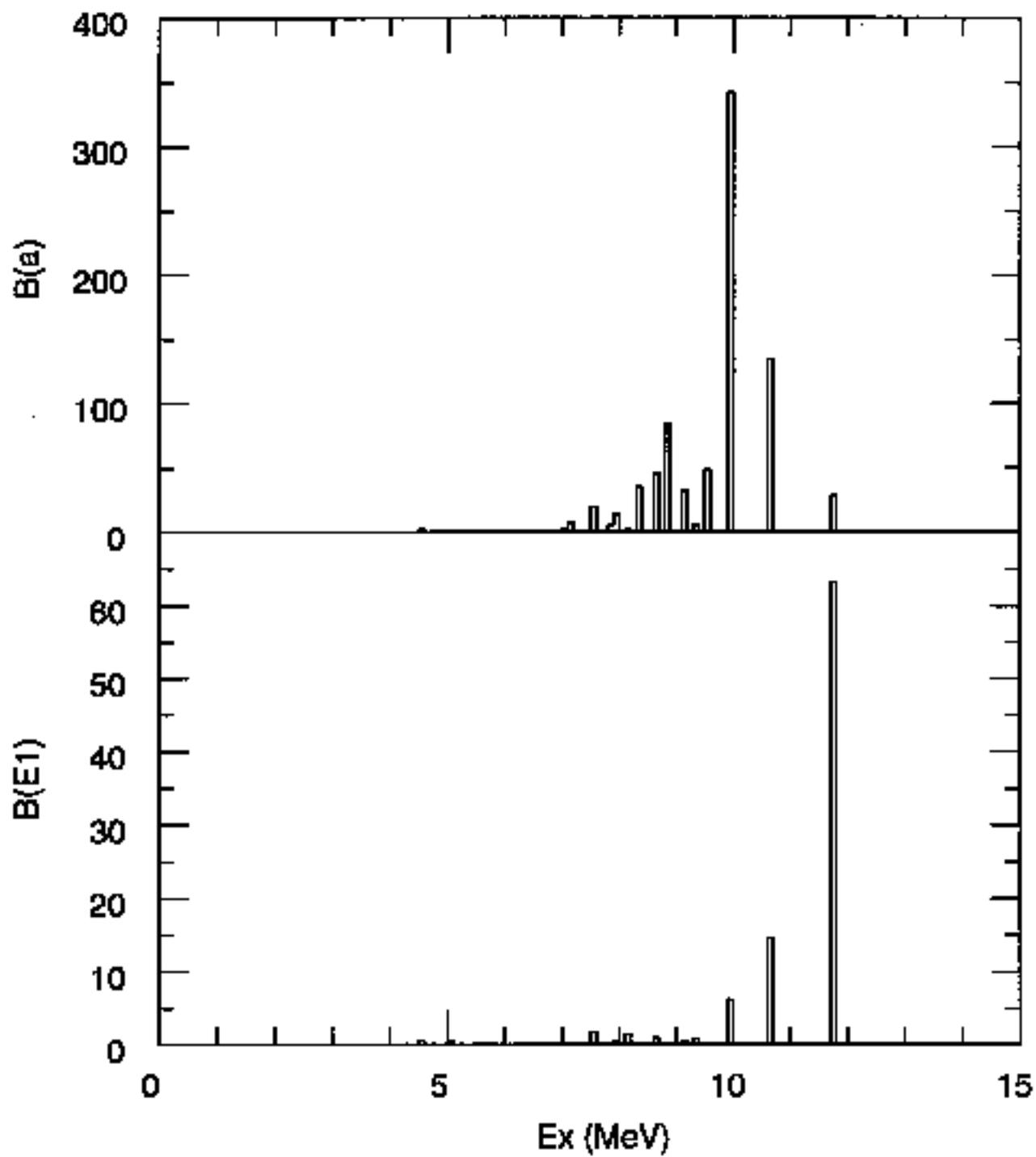
Figure 2: $B(E1)$ and $B(a)$ strength distributions in $^{208}$Pb
obtained with the mixed 1p-1h wave functions.

\newpage
\begin{table}
\caption{Single-particle $\kappa_s$ values for the anapole moments.}
\begin{tabular}{|c|c|c|c|c|c|c|}
Nucleus & single-particle & contribution & $|\Delta E|$ (MeV) & $\kappa_{sp}$ 
& $\kappa_{sn}$ & $\kappa_{s}$ \\
\tableline
$^{207}$Tl  &  $3s_{1/2}$ & WS hole      & 7.63 & 0.042 &       & 0.188 \\
            &             & WS part      & 7.89 & 0.064 &       & 0.287 \\
            &             & WS hole+part &      & 0.106 &       & 0.475 \\
            &             & HF hole      & 8.32 & 0.038 &       & 0.169 \\
            &             & HF part      & 8.67 & 0.058 &       & 0.259 \\
            &             & HF hole+part &      & 0.095 &       & 0.429 \\
\tableline
$^{207}$Pb  &  $3p_{1/2}$ & WS hole      & 7.89 &       & -0.044 & 0.0  \\ 
            &             & WS part      & 6.29 &       & -0.046 & 0.0  \\
            &             & WS hole+part &      &       & -0.090 & 0.0  \\
            &             & HF hole      & 7.83 &       & -0.044 & 0.0  \\
            &             & HF part      & 6.64 &       & -0.044 & 0.0  \\
            &             & HF hole+part &      &       & -0.088 & 0.0  \\
\tableline
$^{209}$Bi  &  $1h_{9/2}$ & WS hole      & 11.44 & 0.062 &       & 0.281 \\
            &             & WS part      & 7.36  & 0.024 &       & 0.106 \\
            &             & WS hole+part &       & 0.086 &       & 0.386 \\
            &             & HF hole      & 11.43 & 0.063 &       & 0.283 \\
            &             & HF part      & 9.10  & 0.019 &       & 0.073 \\
            &             & HF hole+part &       & 0.079 &       & 0.356 \\
\tableline
$^{133}$Sb  &  $1g_{7/2}$ & WS hole      & 12.47 & 0.048 &       & 0.216 \\
            &             & WS part      &  7.87 & 0.023 &       & 0.103 \\
            &             & WS hole+part &       & 0.071 &       & 0.319 \\
            &             & HF hole      & 12.80 & 0.047 &       & 0.210 \\
            &             & HF part      & 9.62  & 0.017 &       & 0.077 \\
            &             & HF hole+part &       & 0.064 &       & 0.287 \\
\end{tabular}
\end{table}

\begin{table}
\caption{Core-polarization corrections for the anapole moments.}
\begin{tabular}{|c|c|c|c|c|c|c|}
Nucleus & single-particle & contribution & $|\Delta E|$ & $\kappa_{sp}$ 
& $\kappa_{sn}$ & $\kappa_{s}$ \\
        &                 &              & (MeV)        & 
&       &       \\
\tableline
$^{207}$Tl 
            & $3s_{1/2}$  & HF part      & 8.67 & 0.058 &       & 0.259 \\
            &             & HF part+CP   & 8.67 & 0.053 & 0.004 & 0.239      \\
\tableline
$^{207}$Pb 
            &  $3p_{1/2}$ & HF part      & 6.64 &       & -0.044 & 0.0  \\
            &         & HF part+CP   & 6.64 & -0.001 & -0.046 & -0.006      \\
\end{tabular}
\end{table}

\newpage
 
\begin{thebibliography}{99}

\bibitem{1.} C.S.~Wood et al.,  Science {\bf 275}, 1759 (1997).

\bibitem{2.} V. V.~Flambaum and D. W.~Murray, Phys. Rev. {\bf
C56}, 1641 (1997).

\bibitem{3.} Ya. B. Zeldovich, Sov. Phys. JETP {\bf 6}, 1184 (1958).


\bibitem{4.} V. V.~Flambaum and I. B.~Khriplovich,
Sov. Phys. JETP {\bf 52}, 835 (1990).

\bibitem{5.} W.C. ~Haxton, Science   
{\bf  275}, 1753 (1997). 


\bibitem{6.} E. G.~Adelberger and W. C. Haxton, Annu. Rev. Nucl.
Part. Sci {\bf 35 }, 501 (1985). 

 
\bibitem{7.} I.~Khriplovich, {\it Parity Nonconservation in Atomic
Phenomena} (Gordon Breach, Philadelphia  1991)


\bibitem{8.} V. F.~Dmitriev and V. B.~Telitsin, Nucl. Phys. 
{\bf A613}, 237 (1997).


\bibitem{9.} N.~Auerbach and A. Klein, Phys. Rev. {\bf C30}, 1032
(1984).

\bibitem{10.} F.~Osterfeld, Rev. of Mod. Phys. {\bf 64}, 491 (1992),
and references therein.


\bibitem{11.} N.~Auerbach, Phys.  Rev. {\bf C45}, R514 (1992). 


\bibitem{12.} A. Bohr and B. R.~Mottelson, 
{\it Nuclear Structure Vol. II}, (W. A. Benjamin, New York, 1975).
 

\bibitem{13.} 
V.F. Dmitriev, J.B. Khriplovich and V.B. Telitsen.
Nucl. Phys. {\bf A577}, 691 (1994).
 
\bibitem{14.} 
A. R. Edmonds, {\it Angular Momentum in Quantum Mechanics},
     (Princeton University Press, 1960).

\bibitem{15.} 
B. Desplanques, J. F. Donoghue and B. R. Holstein, Ann. Phys.
     {\bf 124}, 449 (1980).

\bibitem{16.} 
B. A. Brown, Phys. Rev. {\bf C58}, 220 (1998).

\bibitem{17.} 
B. A. Brown, S. E. Massen, J. I. Escudero, P. E. Hodgson,
     G. Madurga and J. Vinas, J. Phys. {\bf G9}, 423 (1983).

\bibitem{18.} 
J. Streets, B. A. Brown and P. E. Hodgson, J. Phys. {\bf G8}, 839
     (1982).

\bibitem{19.} 
E. K. Warburton and B. A. Brown, Phys. Rev. {\bf C43}, 602 (1991).

\bibitem{20.} 
A. Hosaka, K. I. Kubo and H. Toki, Nucl. Phys. {\bf A244}, 76 (1985).

\bibitem{21.} 
D. H. Glockner and R. D. Lawson, Phys. Lett. {\bf 53B}, 313 (1974).

\bibitem{22.} 
M. Schramm et al., Phys. Rev. {\bf C56}, 1320 (1997); M. Rejmund, 
Dissertation, Univ. of Warsaw, 1999.

\bibitem{23.} 
L. Rydstrom, J. Blomqvist, R. J. Liotta and C. Pomar,
     Nucl. Phys. {\bf A512}, 217 (1990).

\bibitem{24.}
P. A. Vetter, D. N. Meekhof, P. K. Majumder, S. K. Lamoreaux and
E. N. Fortson, Phys. Rev. Lett. {\bf 74}, 2658 (1995).
 

\bibitem{25.} 
G. N. White et al., Nucl. Phys. {\bf A644}, 277 (1998).

\end{thebibliography}
\end{document}